# Lithium Beryllium Phosphide (LiBeP): A possible MgB$_2$-like Superconductor

O. P. Isikaku-Ironkwe[1, 2]
[1]The Center for Superconductivity Technologies (TCST)
Department of Physics,
Michael Okpara University of Agriculture, Umudike (MOUAU),
Umuahia, Abia State, Nigeria
and
[2]RTS Technologies, San Diego, CA 92122

## Abstract

The search for materials similar to magnesium diboride, MgB$_2$, based on structure and electronic similarity did not produce close enough superconductors in terms of Tc. Changing the search to iso-electronic and iso-atomic number equivalents opened new doors to very many possible MgB$_2$-like superconductors. Here we present LiBeP which meets these new conditions. We estimate its Tc to be 34.5K if two-gapped or 17.2K if single gapped.

## Introduction

The discovery of superconductivity at 39K in magnesium diboride [1] presented a model for searching for similar binary, ternary and possibly quaternary systems with low valence electron count and low atomic number like MgB$_2$. The search for MgB$_2$-like superconductors, based on iso-structural and iso-valent similarities failed to find any superconductors [2 - 14] with Tc near to MgB$_2$s, indicating those parameters were not sufficient deciding factors [15 - 17] in superconductivity as previously believed. By looking at other parameters that correlate with superconductivity, such as electronegativity, valence electrons and atomic number, new possible MgB$_2$-like superconductors [17 - 21] were predicted. Using the same model, we herein show that Lithium beryllium phosphide, LiBeP, should be superconducting.

## Properties of LiBeP

Lithium beryllium phosphide, LiBeP, is synthesized from a stoichiometric mixture of Li$_3$P and Be$_3$P$_2$ at a temperature of 660 degrees centigrade. The reaction can be described as:

$$Li_3P + Be_3P_2 = 3LiBeP \qquad (1)$$

The structure of LiBeP has been determined [22, 23] to be tetragonal with a = 3.617 Å and *c* = 6.032 Å. It crystallizes in the P4/nmm, Z =2, in the anti-PbFCl-type structure. A recent computation with Materials Project [24] online resources confirms the spacegroup as P4/nmm No. 129 but with lattice constants of a = 3.606A and c = 6.008A and density 1.99g/cm$^3$.

## MCSD of LiBeP

LiBeP can also be characterized by the material specific characterization dataset (MSCD) scheme [16, 17] as shown in Table 1. Table 1 also has for comparison the MSCDs of $MgB_2$, $LiBSi$, $Be_2Si$ and $Li_2S$. We find that they all share the same atomic number Z and valence electron count, Ne. The $Ne/\sqrt{Z}$ = 0.9847 which implies that it is a superconductor [16]. Thus LiBeP and the others can be classified as $MgB_2$-like materials. A notable feature of the $MgB_2$-like materials is that their formula weights are very close.

## Estimating Tc of LiBeP

The transition temperature Tc of $MgB_2$-like materials can be estimated [16] from

$$T_c = x \frac{Ne}{\sqrt{Z}} K_o \qquad (2)$$

where electronegativity is $x$, valence electron count is Ne, atomic number is Z, and formula weight is Fw. $K_o$ = n(Fw/Z) and n is a number, usually less than 4 and determined empirically for families of superconductors. For $MgB_2$-like superconductors, n = 3.65 for 2-gapped superconductor or half that value for single gap. From equation (2) and Table 1 MSCD, we compute for LiBeP, a Tc of about 34.5K. This computation assumes $MgB_2$'s $K_o$ =22.85. For many ternaries, the Ko is half this value. Thus the Tc (min) of LiBeP may be 17.25K if it is single gapped.

## Discussion

Electronegativity tends to increase with pressure if it has not reached its optimum for a system [25, 26]. We predict that LiBeP will respond to pressure and its Tc will rise to that of MgB$_2$ until its electronegativity reaches 1.7333, the same as MgB$_2$. The MSCDs of LiBSi, Li$_2$S and Be$_2$Si show that their electronegativities are 1.6, 1.5 and 1.6 respectively. That of LiBeP is 1.5333. This implies that their Tcs will be close since their valance electron counts and atomic numbers are the same.

## Conclusion

LiBeP, a tetragonal structured material, meets the symmetry rules required to be a superconductor and also MgB$_2$-like. We computed the Tc to be 17.3K if single gapped and 34.5 K if double gapped.

## Acknowledgements

The author expresses profound gratitude to M J. Schaffer who sponsored these series of researches into MgB$_2$-like superconductivity. M.B. Maple and J. Hirsch at UCSD helped shape some of the ideas herein. J.R. O'Brien at Quantum Design provided useful literature linkages. Earlier enlightening discussions on symmetry with A.O.E. Animalu at the University of Nigeria, have also been continually useful.

## References

1. J. Nagamatsu, N. Nakagawa, T. Muranaka, Y. Zenitani and J. Akimitsu, "Superconductivity at 39K in Magnesium Diboride," Nature 410, 63 (2001)
2. H. Rosner, A. Kitaigorodsky and W.E. Pickett, "Predictions of High Tc Superconductivity in Hole-doped LiBC", Phys Rev. Lett. 88, 127001 (2002).
3. C. Bersier, A. Floris, A. Sanna, G. Profeta, A. Continenza, E.K.U. Gross and "Electronic, dynamical and superconducting properties of CaBeSi", ArXiv: 0803.1044 (2008).
4. Hyoung Jeon Choi, Steven G. Louie and Marvin L. Cohen, "Prediction of superconducting properties of CaB$_2$ using anisotropic Eliashberg Theory", Phys. Rev. B 80, 064503 (2009) and References 1 - 21 in that paper.
5. A. Bharathi, S. Jemima Balaselvi, M. Premila, T. N. Sairam, G. L. N. Reddy, C. S. Sundar, Y. Hariharan "Synthesis and search for superconductivity in LiBC" Arxiv:cond-mat/0207448V1 and references therein., Solid State Comm, (2002), 124, 423
6. Renker, H. Schober, P. Adelmann, P. Schweiss, K.-P. Bohnen, R. Heid ,"LiBC - A prevented superconductor",Cond-mat/0302036

7. A.M. Fogg, J.B. Calridge, G.R. Darling and M.J. Rossiensky "Synthesis and characterization of Li$_x$BC---hole doping does not induce superconductivity". Cond-mat/0304662v1
8. A. Lazicki, C.-S. Yoo, H. Cynn, W. J. Evans, W. E. Pickett, J. Olamit, Kai Liu, and Y. Ohishi, "Search for superconductivity in LiBC at high pressure: Diamond anvil cell experiments and first-principles calculations" Phys. Rev. B 75, 054507 (2007)

9. I. Felner "Absence of superconductivity in BeB$_2$", Physica C 353 (2001) 11 – 13.; D.P. Young, P.W. Adams, J.Y. Chan and F.R. Franczek, "Structure and superconducting properties of BeB$_2$" Cond-mat/0104063
10. B. Lorenz, J. Lenzi, J. Cmaidalka, R.L. Meng, Y.Y. Sun, Y.Y. Xue and C.W. Chu, "Superconductivity in the C32 intermetallic compounds AAl$_{2-x}$Si$_x$, with A=Ca and Sr; and 0.6<$x$<1.2" Physica C, 383, 191 (2002)
11. B. Lorenz, R. L. Meng, and C. W. Chu, "High-pressure study on MgB$_2$", Phys. Rev. B 64, 012507 (2001)
12. R.L. Meng, B. Lorenz, Y.S. Wang, J. Cmaidalka, Y.Y. Xue, J.K. Meen. C.W. Chu "Study of binary and pseudo-binary intermetallic compounds with AlB$_2$ structure" Physica C: 382, 113–116(2002).
13. R.L. Meng, B. Lorenz, J. Cmaidalka, Y.S. Wang, Y.Y. Sun, J. Lenzi, J.K. Meen, Y.Y. Xue and C.W. Chu, "Study of intermetallic compounds isostructural to MgB$_2$, IEEE Trans. Applied Superconductivity, Vol. 13, 3042- 3046 (2002).
14. Cristina Buzea, Tsutomu Yamashita, "Review of superconducting properties of MgB$_2$", Superconductor Science & Technology, Vol. 14, No. 11 (2001) R115-R146
15. O. Paul Isikaku-Ironkwe, "Search for Magnesium Diboride-like Binary Superconductors" http://meetings.aps.org/link/BAPS.2008.MAR.K1.7
16. O. Paul Isikaku-Ironkwe, "Transition Temperatures of Superconductors estimated from Periodic Table Properties", Arxiv: 1204.0233 (2012)
17. O. P. Isikaku-Ironkwe, "Possible High-Tc Superconductivity in LiMgN: A MgB$_2$-like Material", Arxiv: 1204.5389 (2012)
18. O. P. Isikaku-Ironkwe, "Prospects for Superconductivity in LiBSi", Arxiv: 1205.2407 (2012)
19. O. P. Isikaku-Ironkwe, "Is Lithium Sulfide a MgB$_2$-like Superconductor?", Arxiv: 1205.4051 (2012)
20. O. P. Isikaku-Ironkwe, "Beryllium silicide clusters, Be$_{2n}$Si$_n$ (n = 1 – 4) and possible MgB$_2$-like Superconductivity in some of them", Arxiv: 1205.5931 (2012)
21. O. P. Isikaku-Ironkwe, "MgBeC: A potential MgB$_2$-like Superconductor", Arxiv: 1205.6237(2012)
22. Abdallah El Maslout, Jean-Pierre Motte, Alain Courtois, Charles Gleitzer Phosphures Ternaires de Lithium Li$_{2n-3}$M$^n$P$_{n-1}$ (M = Be, Cd, Sn) de Structure Antifluorine ou Derivee J. Solid State Chem., 7), 250-254 (1973)

23. Abdallah El Maslout, Jean-Pierre Motte, Alain Courtois, Charles Gleitzer
    Phosphures Ternaires de Lithium II. Structure cristalline de LiBeP
    *J. Solid State Chem.*, *Volume 15(3), 213-217 (1975)*
24. Materials Project, "Material 9915: LiBeP" www.materialsproject.org/tasks/9915 ; accessed June 8, 2012
25. R. Asokanami and R. Manjula, "Correlation between electronegativity and superconductivity", Phys. Rev. B 39 4217 – 4221, (1989)
26. W. Gordy, " A new method of determining electronegativity from other atomic properties", Phy. Rev. 69, 604 -607, (1946)

## Table

| | Material | $\mathcal{X}$ | Ne | Z | Ne/$\sqrt{Z}$ | Fw | Fw/Z | Tc(K) | Ko |
|---|---|---|---|---|---|---|---|---|---|
| 1 | $MgB_2$ | 1.7333 | 2.667 | 7.3333 | 0.9847 | 45.93 | 6.263 | 39 | 22.85 |
| 2 | LiBeP | 1.5333 | 2.667 | 7.3333 | 0.9847 | 46.92 | 6.40 | 34.5 | 22.85 |
| 3 | LiBSi | 1.6 | 2.667 | 7.3333 | 0.9847 | 45.84 | 6.25 | 36 | 22.85 |
| 4 | $Be_2Si$ | 1.6 | 2.667 | 7.3333 | 0.9847 | 46.11 | 6.29 | 36 | 22.85 |
| 5 | $Li_2S$ | 1.5 | 2.667 | 7.3333 | 0.9847 | 45.95 | 6.27 | 33.8 | 22.85 |

Table 1: MSCD of $MgB_2$, LiBeP, LiBSi, $Be_2Si$ and $Li_2S$. Ne and Z is the same for all, hence Ne/$\sqrt{Z}$ is also the same. Formula weight (Fw) are very close and differ by less than 1.0